\documentclass{ws-procs961x669}            
\usepackage[per-mode=fraction,separate-uncertainty=true]{siunitx}
\usepackage{color}

\DeclareSIUnit{\sqrthz}{\ensuremath{\sqrt{\text{\hertz}}}}
\DeclareSIUnit{\mnoise}{\metre\per\sqrthz}

\begin{document}
\title{LISA Pathfinder}

\author{
	M~Armano$^a$,
	H~Audley$^b$,
	J~Baird$^c$,
	P~Binetruy $^{d,}$\thanks{Deceased 30 September 2012},
	M~Born$^b$,
	D~Bortoluzzi$^e$,
	E~Castelli$^f$, 
	A~Cavalleri$^f$,
	A~Cesarini$^g$,
	A\,M~Cruise$^h$,
	K~Danzmann$^b $,
	M~de Deus Silva$^a$,
	I~Diepholz$^b$,
	G~Dixon$^h$,
	R~Dolesi$^f$,
	L~Ferraioli$^i$,
	V~Ferroni$^f$,
	E\,D~Fitzsimons$^j$,
	M~Freschi$^a$,
	L~Gesa$^k$,
	F~Gibert$^f$,
	D~Giardini$^i$,
	R~Giusteri$^f$,
	C~Grimani$^g$,
	J~Grzymisch$^l$,
	I~Harrison$^m$,
	G~Heinzel$^b$,
	M~Hewitson$^b$,
	D~Hollington$^c$,
	D~Hoyland$^h$,
	M~Hueller$^f$,
	H~Inchausp\'e$^d$, 
	O~Jennrich$^l$,
	P~Jetzer$^n$,
	N~Karnesis$^b$,
	B~Kaune$^b$,
	N~Korsakova$^o$,
	C\,J~Killow$^o$,
	J\,A~Lobo$^{k,}$\thanks{Deceased 30 September 2012},
	I~Lloro$^k$,
	L~Liu$^f$,
	J\,P~L\'opez-Zaragoza$^k$,
	R~Maarschalkerweerd$^m$,
	D~Mance$^i$
	N~Meshskar$^i$,
	V~Mart\'{i}n$^k$,
	L~Martin-Polo$^a$,
	J~Martino$^d$,
	F~Martin-Porqueras$^a$,
	I~Mateos$^k$,
	P\,W~McNamara$^l$,
	J~Mendes$^m$,
	L~Mendes$^a$,
	M~Nofrarias$^k$,
	S~Paczkowski$^b$$^{\star}$,
	M~Perreur-Lloyd$^o$,
	A~Petiteau$^d$,
	P~Pivato$^f$,
	E~Plagnol$^d$,
	J~Ramos-Castro$^p$,
	J~Reiche$^b$,
	D\,I~Robertson$^o$,
	F~Rivas$^k$,
	G~Russano$^f$,
	J~Slutsky$^q$,
	C\,F~Sopuerta$^k$,
	T~Sumner$^c$,
	D~Texier$^a$,
	J\,I~Thorpe$^q$,
	D~Vetrugno$^f$,
	S~Vitale$^f$,
	G~Wanner$^b$,
	H~Ward$^o$,
	P~Wass$^{c,r}$,
	W\,J~Weber$^f$,
	L~Wissel$^b$,
	A~Wittchen$^b$,
	and P~Zweifel$^i$
}

\address{$^a$ European Space Astronomy Centre, European Space Agency, Villanueva de la
	Ca\~{n}ada, 28692 Madrid, Spain}
\address{$^b$Albert-Einstein-Institut, Max-Planck-Institut f\"ur Gravitationsphysik und Leibniz Universit\"at Hannover,
	Callinstra{\ss}e 38, 30167 Hannover, Germany \\ $^{\star}$E-mail: sarah.paczkowski@aei.mpg.de}
\address{$^c$High Energy Physics Group, Physics Department, Imperial College London, Blackett Laboratory, Prince Consort Road, London, SW7 2BW, UK }
\address{$^d$APC, Univ Paris Diderot, CNRS/IN2P3, CEA/lrfu, Obs de Paris, Sorbonne Paris Cit\'e, France * Deceased 30 March 2017}
\address{$^e$Department of Industrial Engineering, University of Trento, via Sommarive 9, 38123 Trento, 
	and Trento Institute for Fundamental Physics and Application / INFN}
\address{$^f$Dipartimento di Fisica, Universit\`a di Trento and Trento Institute for 
	Fundamental Physics and Application / INFN, 38123 Povo, Trento, Italy}
\address{$^g$DISPEA, Universit\`a di Urbino ``Carlo Bo'', Via S. Chiara, 27 61029 Urbino/INFN, Italy}
\address{$^h$The School of Physics and Astronomy, University of
	Birmingham, Birmingham, UK}
\address{$^i$Institut f\"ur Geophysik, ETH Z\"urich, Sonneggstrasse 5, CH-8092, Z\"urich, Switzerland}
\address{$^j$The UK Astronomy Technology Centre, Royal Observatory, Edinburgh, Blackford Hill, Edinburgh, EH9 3HJ, UK}
\address{$^k$Institut de Ci\`encies de l'Espai (CSIC-IEEC), Campus UAB, Carrer de Can Magrans s/n, 08193 Cerdanyola del Vall\`es, Spain $\dag$ Deceased 30 September 2012}
\address{$^l$European Space Technology Centre, European Space Agency, 
	Keplerlaan 1, 2200 AG Noordwijk, The Netherlands}
\address{$^m$European Space Operations Centre, European Space Agency, 64293 Darmstadt, Germany }
\address{$^n$Physik Institut, 
	Universit\"at Z\"urich, Winterthurerstrasse 190, CH-8057 Z\"urich, Switzerland}
\address{$^o$SUPA, Institute for Gravitational Research, School of Physics and Astronomy, University of Glasgow, Glasgow, G12 8QQ, UK}
\address{$^p$Department d'Enginyeria Electr\`onica, Universitat Polit\`ecnica de Catalunya,  08034 Barcelona, Spain}
\address{$^q$Gravitational Astrophysics Lab, NASA Goddard Space Flight Center, 8800 Greenbelt Road, Greenbelt, MD 20771 USA}
\address{$^r$Department of Physics, University of Florida, 2001 Museum Rd, Gainesville, FL 32603, USA}

\begin{abstract}
Since the 2017 Nobel Prize in Physics was awarded for the observation of gravitational waves, it is fair to say that the epoch of gravitational wave astronomy (GWs) has begun. However, a number of interesting sources of GWs can only be observed from space.
To demonstrate the feasibility of the Laser Interferometer Space Antenna (LISA), a future gravitational wave observatory in space, the LISA Pathfinder satellite was launched on December, 3rd 2015. Measurements of the spurious forces accelerating an otherwise free-falling test mass, and detailed investigations of the individual subsystems needed to achieve the free-fall, have been conducted throughout the mission. This overview article starts with the purpose and aim of the mission, explains satellite hardware and mission operations and ends with a summary of selected important results and an outlook towards LISA.
From the LISA Pathfinder experience, we can conclude that the proposed LISA mission is feasible.
\end{abstract}
\keywords{Gravitational Waves, Interferometers, Space Research Instruments, Laser Metrological applications} 
\bodymatter
\section{Introduction to the LISA Pathfinder project}
LISA Pathfinder is a technology demonstrator mission for the Laser Interferometer Space Antenna, LISA. To understand the necessity and the main goal of LISA Pathfinder, let us review some important aspects of LISA.
\vspace{-0.2cm}
\subsection{The Laser Interferometer Space Antenna LISA}
\subsubsection{Short summary of gravitational wave sources in the LISA band}
LISA is a mission concept for a future gravitational wave observatory in space. It will not be competing with the ground-breaking discoveries of the LIGO-Virgo collaboration but opens the opportunity to observe gravitational waves at lower frequencies. To be more precise, with LISA we aim to be able to measure gravitational waves in the frequency regime from $\SI{20}{\micro\Hz}$ to $\SI{1}{\Hz}$. In this measurement band, we aim to observe gravitational waves from several very interesting sources, which are for example supermassive black hole binaries and extreme mass ratio inspirals (EMRIs). 

As explained in \citenum{the_gravitational_universe}, LISA is a great instrument to observe supermassive black hole binaries, especially for those with large redshifts of $z \approx 10$ and beyond with relatively small masses between $10^4 M_\odot$ and $10^7 M_\odot$. Black holes at these redshifts are difficult to observe using electromagnetic radiation in the optical regime because it is suppressed for certain frequencies and for redshifts larger than approximately $6$. Also X-Ray observations may suffer from a deterioration of the signals of these sources due to crowded sources and unresolved background light, difficulties that do not apply for LISA. These supermassive black hole binaries are very important to discover the formation of seed black holes around the cosmic dawn.  LISA is expected to enable us to study the black hole binaries with masses between $10^4 M_\odot$ and $10^7 M_\odot$ out to redshifts of $20$, if they exist. In general, LISA can also be seen as a large black hole binary search over a vast range of masses and redshifts in nearly all directions which is, even with future observatories, only in part accessible in the electromagnetic regime. In total, a coalescence rate between 10 and 100 per year is expected to be observable with LISA. The observation of coalescences are especially interesting because they will allow us to better understand the accretion mechanisms of black holes.

EMRIs consist of a compact object, which, in the case of LISA signals, is more likely a stellar mass black hole than a neutron star, which is in a highly relativistic orbit around a massive black hole. As such, EMRIs are an excellent opportunity to study gravity in the strong-field and non-linear regime. In addition, they provide another opportunity to study the stellar dynamics around a massive black hole. This information, as well as further details, can be found in \citenum{the_gravitational_universe}.

Furthermore, LISA will be sensitive to a stochastic background of gravitational waves with
an energy density of $\Omega_{\mathrm{GW}} \simeq 10^{-13}$ or higher between between $\SI{1}{\milli\Hz}$ and $\SI{10}{\milli\Hz}$\cite{privateCommunicationGermano}. Conservative estimates for the previous eLISA concept required an energy density of $\Omega_{\mathrm{GW}} \simeq 10^{-10}$ in the present universe\cite{privateCommunicationChiara}. Interestingly, signals with such a feature can be originated by both astrophysical and cosmological sources\cite{privateCommunicationGermano}. For instance, from the detection of GW150914, the stochastic background from binary black holes is predicted to have an energy density $\Omega_{\mathrm{GW}} = 1.1^{+2.7}_{-0.9}\cdot 10^{-9}$ at $\SI{25}{\Hz}$\cite{stoch_backgrd_imp_from_GW150914}. Concerning the cosmological sources, the most studied are those related to the inflationary processes, topological defects and first-order phase transition\cite{LISA_probing_inflation_with_GWs}\cite{LISA_GWs_from_cosmological_phase_transitions}. To create a visible signal in LISA, at the time of production in the radiation dominated era in the early universe, a fraction of $\Omega_{\mathrm{GW}} > 10^{-5}$ of the energy density of the universe must have been converted into gravitational radiation\cite{the_gravitational_universe}. All these processes involve physics beyond the standard model of particle physics with a new physics scale from the electroweak up to the Planck mass scale. Due to the sensitivity to these scales, LISA is complementary to particle physics experiments as e.g. the LHC\cite{privateCommunicationGermano}.
 
Moreover, with LISA, we expect to be able to predict the time to coalescence of black-hole binaries which pass from the LISA measurement band to the advanced LIGO frequency range within the LISA lifetime, with an accuracy of approximately $\SI{10}{\second}$ and their sky location with an accuracy of one square degree\cite{Alberto_Observation_of_LIGO_sources_in_LISA}.

Finally, it should be mentioned that the list of possible sources is by far not complete. Especially, the unpredicted sources which might be measured could open a whole new discovery space.
\subsubsection{LISA is only possible in Space}
The main reason why LISA is only possible in space is that towards lower frequencies, in a ground-based gravitational wave observatory such as LIGO, the seismic and gravity gradient noise becomes limiting. Gravity gradient noise is also known as Newtonian gravity noise and describes the disturbances caused by changes in local Newtonian gravity due to moving objects close to the extremely sensitive detector \cite{Saulson1994}.
In addition, it is difficult to achieve sufficient thermal and mechanical stability for the corresponding long measurement durations. To some extent, achieving the required thermal stability is also non-trivial in space. Moreover, in space, it is much easier to have even longer arms to increase the sensitivity of the instrument. 

The gravitational wave observatory LISA, as described in \citenum{LISA_L3}, is characterised by the triangular constellation which is formed by three satellites. This constellation is trailing approximately $\SI{20}{\degree}$ behind the Earth on a heliocentric orbit. These orbits also introduce a `cartwheel' rotation of the whole constellation but at the same time they keep the relative separation of the satellites as well as the angles of the triangle as constant as possible. Laser light is exchanged between the satellites, which are planned to be $2.5$ million $\SI{}{\kilo\metre}$ away from each other. The sides of the triangle correspond to the arms of a ground-based gravitational wave observatory and hence, they are also called arms. Each satellite hosts two test masses which define the beginning and the end of each of the arms. Gravitational waves now cause tiny variations in the relative distance of the two such test masses. These are measured using heterodyne laser interferometry. Therefore, there is an optical bench for each test mass, so two on each satellite. The measurement of the relative distance in between two test masses on two different satellites is split into three measurements: a local test mass to satellite measurement, a satellite to satellite measurement and another local test mass to satellite measurement on the far satellite. Accordingly, there will be local heterodyne interferometry and inter-satellite heterodyne interferometry required. In addition, the laser light of the two different optical benches on the same satellite needs to be compared to enable the laser frequency noise suppression via the Time-Delay Interferometry (TDI) algorithm\cite{Otto2012}. Therefore, in the current design, an optical fibre is linking the two optical benches on a LISA satellite.
\subsubsection{LISA requires quiet test masses}
Most importantly, to measure gravitational waves, LISA requires free-falling test masses. This means they are subject to no other force than gravity. To put it simply, we have to ensure that a measured signal is due to the effect of gravitational waves and not caused by interaction between any of the two test masses and the respective satellite or its components. The residual acceleration of an otherwise free-falling test mass, originating from a number of possibly unknown, spurious forces caused by the satellite and the satellite environment is what we call $\Delta g$. It is a noisy signal and the smaller the noise level, the less residual acceleration is present and the quieter and closer to free-fall is the test mass.

Just how very close to a perfect free-fall, each pair of test masses has to be can be seen from the requirement \cite{LISA_L3}
\begin{equation}
\mathrm{S^{\frac{1}{2}}}_{g, \mathrm{LISA}} \leq  3 \cdot 10^{-15} \SI{}{\metre\per\second\squared\per\sqrthz} \sqrt{1+\left(\frac{\SI{0.4}{\milli\Hz}}{f}\right)^2}\sqrt{1 + \left(\frac{f}{\SI{8}{\milli\Hz}}\right)^4}
\end{equation}
that applies to the square root of the power spectral density of the residual acceleration of a single test mass $g$. Ideally, these fluctuations would vanish but this is not possible in reality. If the fluctuations are below the requirement, the sensitivity of LISA is good enough to answer the science questions.  However, it is impossible to test the technology for LISA at these tiny acceleration levels in a standard laboratory on Earth. A torsion pendulum facility, however, gets close to the required accuracy but is then limited by mechanical thermal noise at lower frequencies and readout noise towards higher frequencies\cite{torsion_pendulum_UTN_2002}. Let us note furthermore that to measure $g$ at $\SI{}{\milli\Hz}$ frequencies, it is necessary to measure $\approx \SI{17}{\minute}$. That is why a drop-tower would not be sufficient. In addition, this requirement is more stringent than on previous drag-free missions. For example on GOCE, the residual acceleration requirement is relaxed by two orders of magnitude in comparison to LISA Pathfinder \cite{GOCE}. That is why a test in space, as it is done with the LISA Pathfinder Mission, is required. 

\subsection{LISA Pathfinder Mission Goal}
Accordingly, the LISA Pathfinder mission goal is to demonstrate the technology for LISA, which means foremost to show that a nearly perfect free-fall is feasible. In comparison to LISA, the residual differential acceleration requirement is relaxed by one order of magnitude to\cite{2016PhRvL.116w1101A} 
\begin{equation}
\mathrm{S^{\frac{1}{2}}}_{\Delta g, \mathrm{LPF}} \leq 30 \cdot 10^{-15} \SI{}{\metre\per\second\squared\per\sqrthz} \sqrt{1+\left(\frac{f}{\SI{3}{\milli\Hz}}\right)^4}  
\label{eq:LPF_req}
\end{equation}
and restricted to the frequency range from $\SI{1}{\milli\Hz}$ to $\SI{30}{\milli\Hz}$. In contrast to LISA, LISA Pathfinder is not designed to observe gravitational waves because the distance in between the two test masses, as explained below, is too short.  

\section{The LISA Pathfinder satellite}
\subsection{LISA Pathfinder hosts two free-floating test masses}
A first and idealised idea to prove Equation \ref{eq:LPF_req} could be to put a single test mass carefully into a quiet environment in space. However, to measure an acceleration, a reference point and a measurement system is needed. Thus, we have a satellite that hosts the measurement system and shields the test mass from, for example, solar radiation pressure. In theory, already the satellite could be used as a reference point for the acceleration measurement. However, this reference point itself is too noisy for the acceleration levels we want to measure on LISA Pathfinder. Therefore, we also have a quiet, free-floating reference mass. 

These two free-falling test masses are at the core of LISA Pathfinder. Free-falling means, they are not physically connected or touched by anything. The test masses are quasi cubes made of a gold-platinum alloy and their edges are $\SI{46 \pm 0.005}{\milli\metre}$ long\cite{2016PhRvL.116w1101A}. The material has been chosen to have a low magnetic susceptibility of $\chi_{\mathrm{M}} \approx 10^{-5}$ combined with a high density\cite{LPF_mission_status_2011}. Each test mass weighs $\SI{1.92 \pm 0.001}{\kilogram}$. A comparatively large mass minimises undesired gravitational interaction with the surroundings. They are $\SI{376.00 \pm 0.05}{\milli\metre}$ apart from each other, as can be seen in Figure \ref{fig:key_components_to_achieve_free_fall}. The reason they are not perfect cubes is the caging and release mechanism. It has to fix the test masses during launch because a loose test mass during the rocket launch would cause severe damage. In addition, it has to release the test masses smoothly in the final orbit. 
The main measurement on LISA Pathfinder is now the change in relative distance in between the two free-falling test masses. The relative residual acceleration is then obtained as the second derivative of this measurement. The details will be given in Section \ref{subsec:delta_g_explained}.

\subsection{The LISA Pathfinder Drag-Free and Attitude Control System}
\begin{figure}
	\centering
	\includegraphics[width=0.4\textwidth]{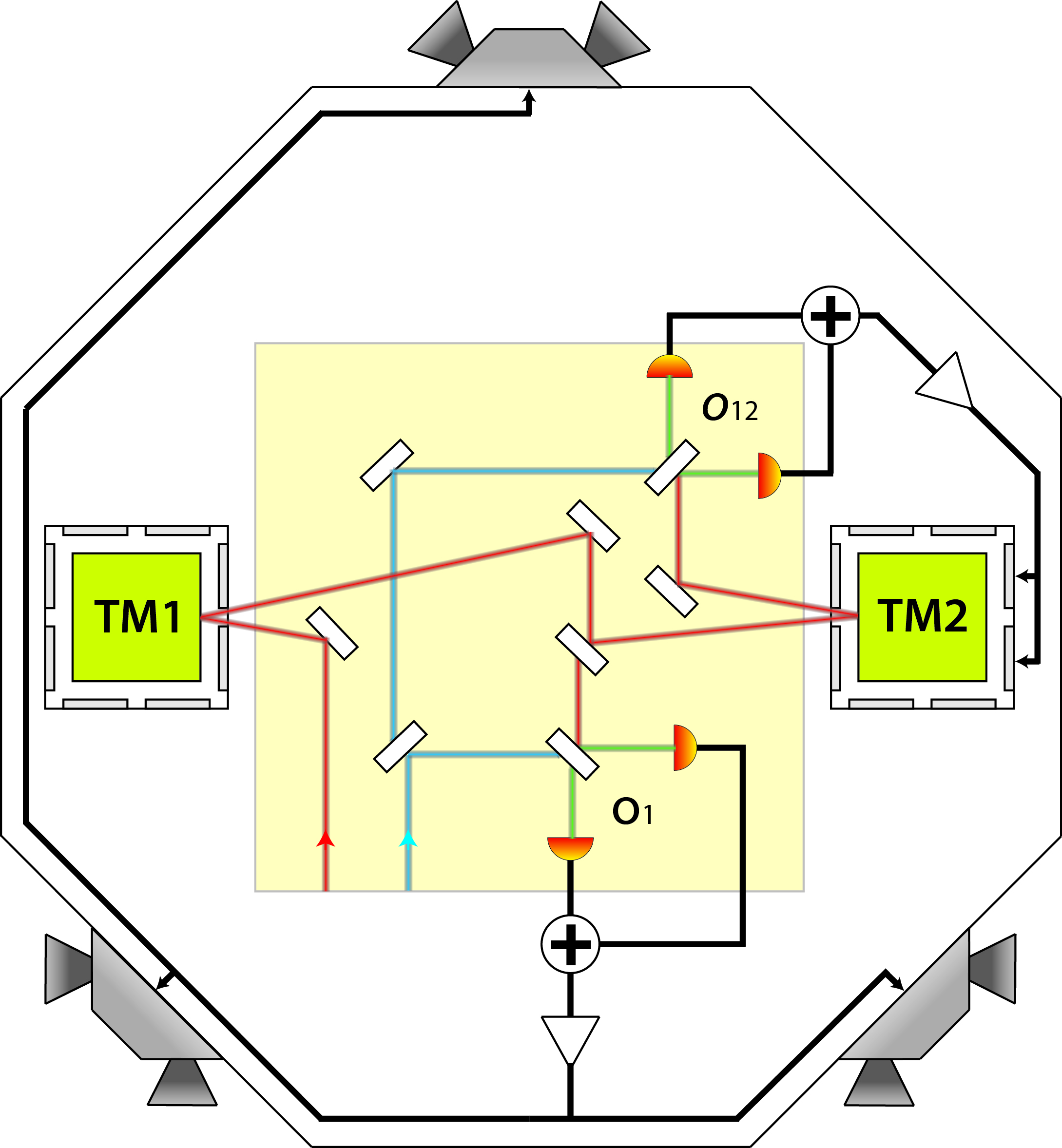}
	\caption{A schematic representation of the LISA Pathfinder components used by the Drag-Free and Attitude Control System (DFACS). $o1$ is the measurement of the interferometer X1 and $o12$ is the measurement of the interferometer X12. Reprint from \citenum{LPF_overview_LISA_Symp_Florida}.}
	\label{fig:LPF_control_diagram}
\end{figure}

On LISA Pathfinder, a Drag-Free and Attitude Control System (DFACS) is being used to minimise the undesired influence of the satellite on the free-floating test masses. Our main science mode, which is used for our measurements of the residual acceleration, works like this: Along the sensitive axis, which is defined by the line that connects the two test masses, one of the test masses, usually called TM1, is not subject to any control force. The satellite is then controlled by the DFACS to follow the motion of TM1. The DFACS system obtains the position of the TM1 with respect to the satellite, as measured by the interferometer X1. The output of this interferometer is called o1, as can be seen in Figure \ref{fig:LPF_control_diagram}. From this, it determines the necessary thrust that has to be applied by the thrusters to have the satellite follow TM1. This control loop is called the drag-free loop. Also the other test mass is made to follow the motion of TM1 in this mode. The measurement of the relative position of the two test masses, as measured with the interferometer X12, is called o12. Using this o12 measurement, the DFACS system also estimates the force that has to be applied to the other test mass such that it follows TM1. This loop is called the suspension loop. The necessity for the suspension can be understood by thinking of the two test masses moving in opposite directions along the sensitive axis. Then the satellite would not be able to follow them both. So one of them, usually TM2, is made to follow the other, usually TM1. However, the commanded force is known to the data analysis. It is used to estimate the true applied forces onto TM2 such that they can be subtracted in post-processing. It is also important to mention that the unity gain frequency of this so-called suspension loop is near the end of our measurement band. This minimises the impact of the control forces on the measurement of the residual acceleration noise. More details can be found in Section \ref{subsec:delta_g_explained}. In addition, a dedicated control mode is implemented on LISA Pathfinder which replaces the continuous control of TM2 along x with an intermittent control scheme. This experiment is called the drift mode or free-flight experiment (see Section \ref{subsec:GRS_measurements_and_dm}). 
\subsection{Key subsystems to achieve free-fall on LISA Pathfinder}
\begin{figure}
	\centering
	\includegraphics[width=0.5\textwidth]{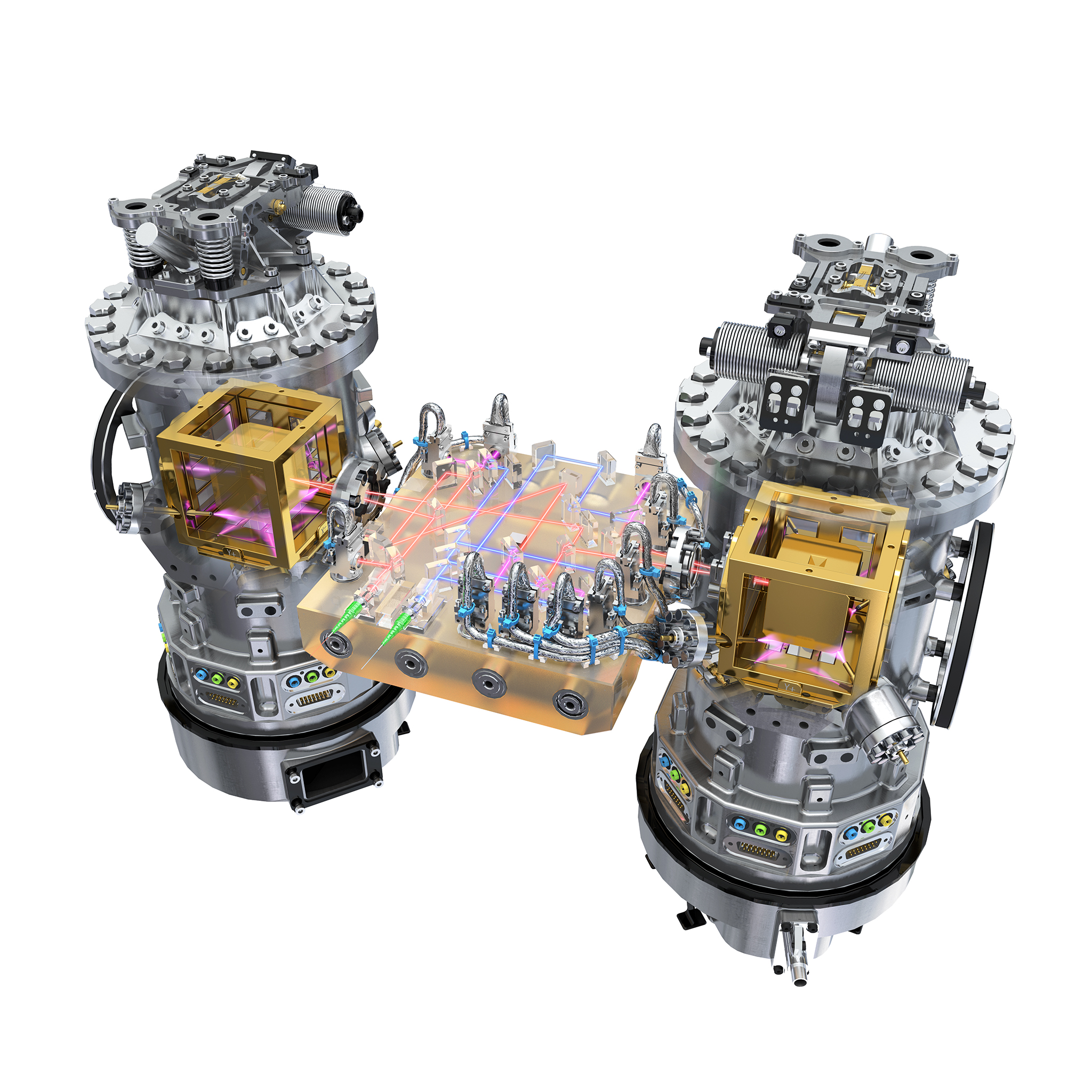}
	\caption{The core of LISA Pathfinder. The two golden cubes are the test masses inside their respective electrode housings inside a vacuum tank. The optical bench is discernible in the centre. Image: ESA/ATG medialab.}
	\label{fig:key_components_to_achieve_free_fall}
\end{figure}
To achieve the necessary levels of free-fall, not only the DFACS as explained above is essential but it can work only together with a number of key subsystems, as shown in Figure \ref{fig:key_components_to_achieve_free_fall}. That means also that we test all these systems for LISA. A key subsystem is the Gravity Reference Sensor (GRS). This system includes the two test masses as well as the electrodes mounted on the electrode housing and the corresponding front-end electronics (FEE). Each of the two test masses is in a vacuum tank. They have an optical window such that the light can reach the test masses for the measurement of the relative position in between the two. Another key component is the discharge mechanism. In contrast to other space missions, for example MICROSCOPE \cite{MICROSCOPE_status_2015}, where the discharging is done via a wire, the test masses on LISA Pathfinder are not in contact with anything and thus have no discharge wire. Instead, a UV lamp is used for discharging via the photoelectric effect. In addition, thrusters that allow the satellite to follow the test mass, while producing minimal undesired noise, are needed. These are the cold-gas $\SI{}{\micro\N}$ thrusters, which are also used on the MICROSCOPE and GAIA mission\cite{EP_thrusters_LISA_symp_Florida}. It is also important to mention that the temperature on LISA Pathfinder has to be stable and that the magnetic fields on board have to be minimised to achieve the required level of free-fall. Another key component is the Optical Metrology System (OMS).  In Figure \ref{fig:key_components_to_achieve_free_fall}, the OMS is easily discernible via the optical bench that is located in the centre of the picture. The optical bench is made out of a material with ultra-low thermal expansion, which is called $\text{Zerodur}^\text{\copyright}$. The lines in red and blue mark the paths of the laser light through the mirrors and beam splitters. 
\subsection{The LISA Pathfinder Optical Metrology System}
The main purpose of the optical metrology system is to measure the distance of the free-falling test mass with respect to the quiet reference test mass with a required precision of\cite{oldSci_Req}
\begin{equation}
	\mathrm{S^{\frac{1}{2}}}_{\delta x}  
	\leq \SI{9}{\pico\mnoise}
	\sqrt{ 1+\left(\frac{\SI{3}{\milli\Hz}}{f}\right)^{4}}\,.
\end{equation}
Here, $\mathrm{S^{\frac{1}{2}}}_{\delta x}$ is the square root of the power spectral density of the measured fluctuations. To achieve this precision, a heterodyne laser interferometry set-up has been chosen. A Nd:YAG laser with a wavelength of $\SI{1064}{\nano\meter}$ and an output power of a few tens of $\SI{}{\milli\watt}$ is located in the so called reference laser unit (not discernible in Figure \ref{fig:key_components_to_achieve_free_fall}). From there, the light travels to the laser modulator unit where the light is split into two beams. Each of them is frequency shifted. Then the two beams travel via optical fibres to the optical bench. They leave the fibres via the fiber injector optical sub assembly which can be recognized by the green ends in the centre of Figure \ref{fig:key_components_to_achieve_free_fall}. On the optical bench, the two beams are brought to interference again. The resulting beat note of the light is recorded by the photodiodes. These are the metallic pieces on the optical bench in Figure \ref{fig:key_components_to_achieve_free_fall} whose cables are held by blue cable ties. 
The phase of this signal is the measured quantity which contains, for example, the information on the relative distance in between the two test masses. However, on the optical bench, the two beams are brought to interference not only once but in four different ways. Hence, we have four different interferometers, as shown in Figure \ref{fig:4_IFOs}.\def\figsubcap#1{\par\noindent\centering\footnotesize(#1)}
\begin{figure}[h]%
	\begin{center}
		\parbox{2.1in}{\includegraphics[width=2in]{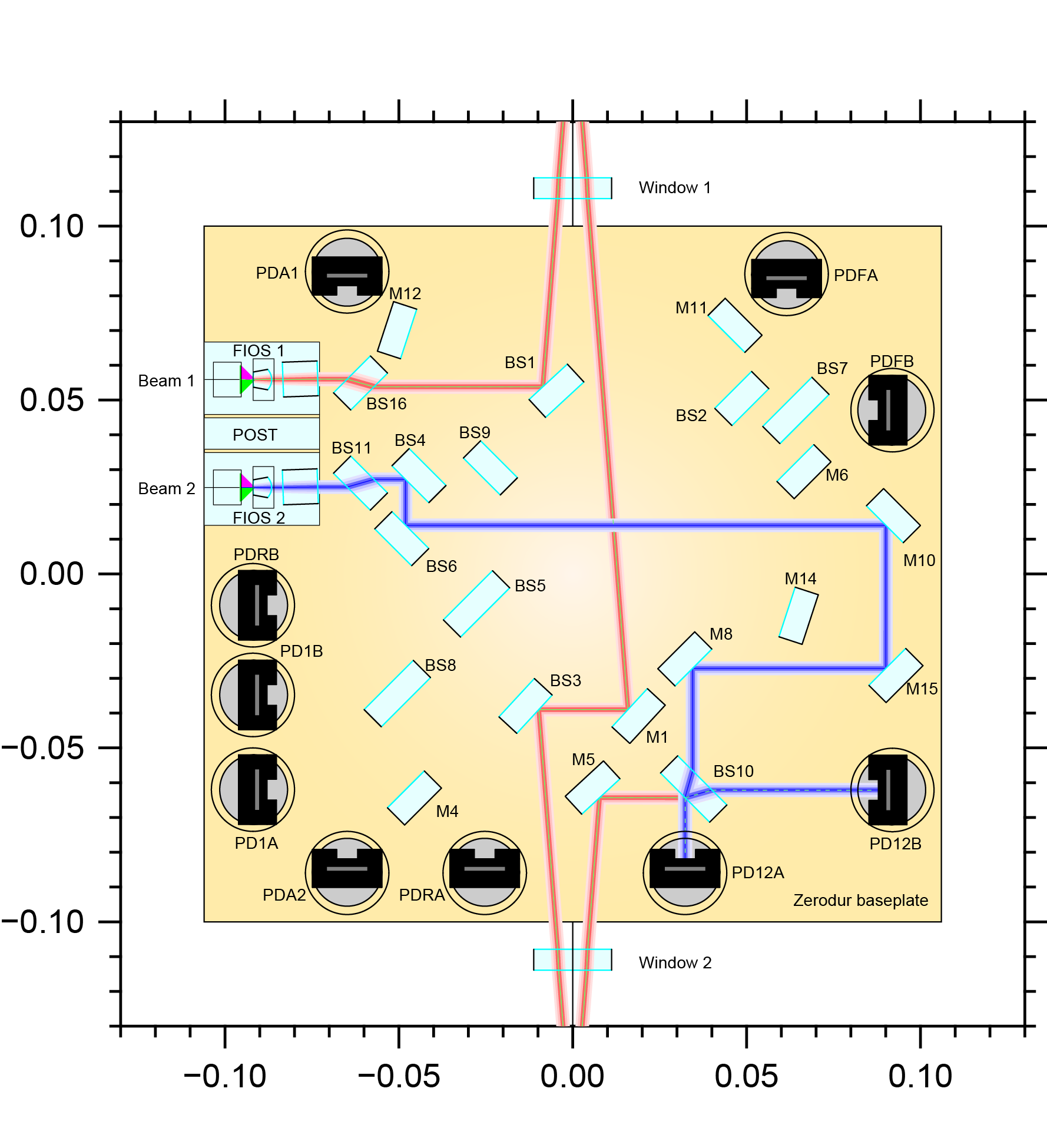}\figsubcap{a}}
		\hspace*{4pt}
		\parbox{2.1in}{\includegraphics[width=2in]{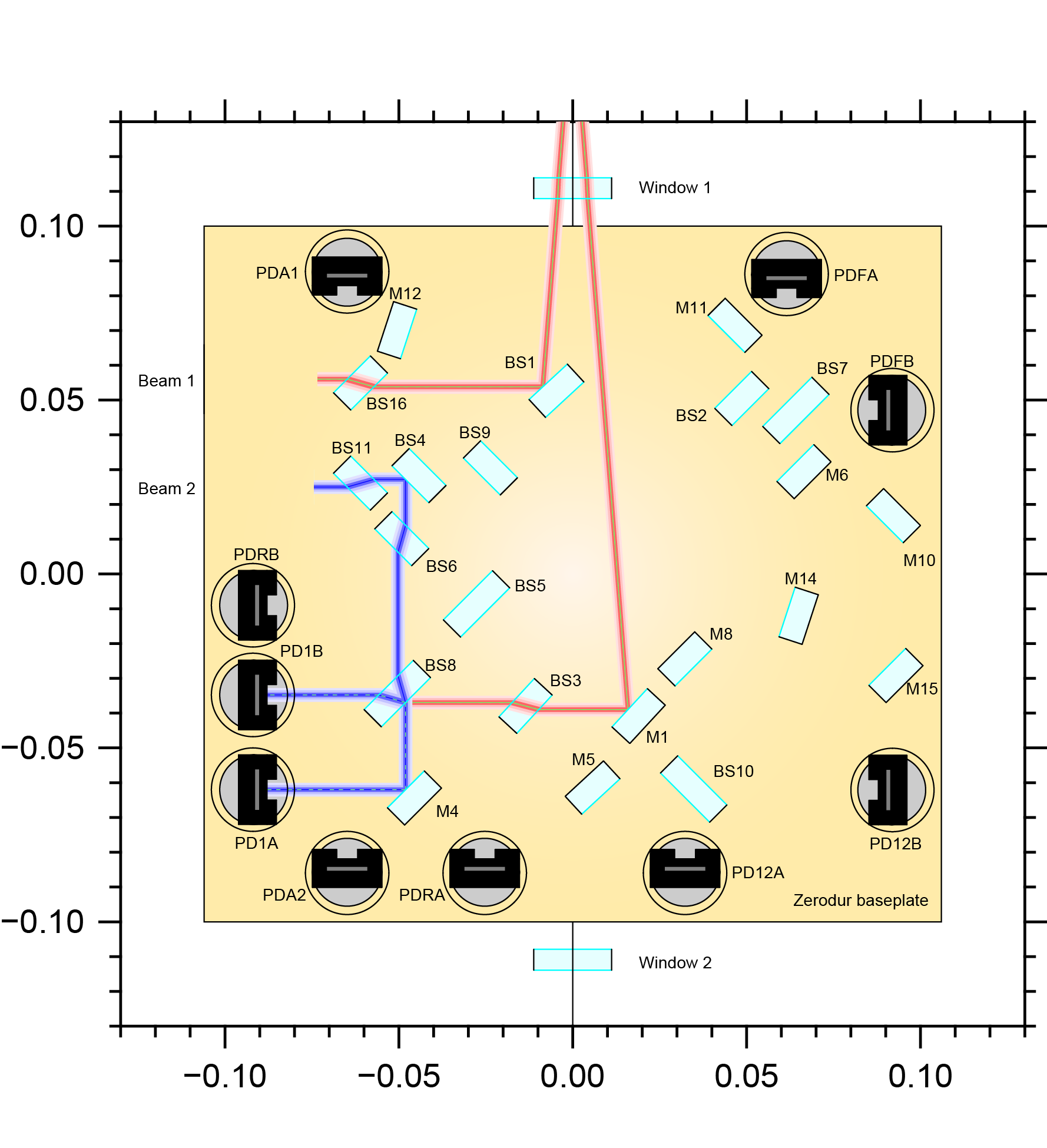}\figsubcap{b}}
		\parbox{2.1in}{\includegraphics[width=2in]{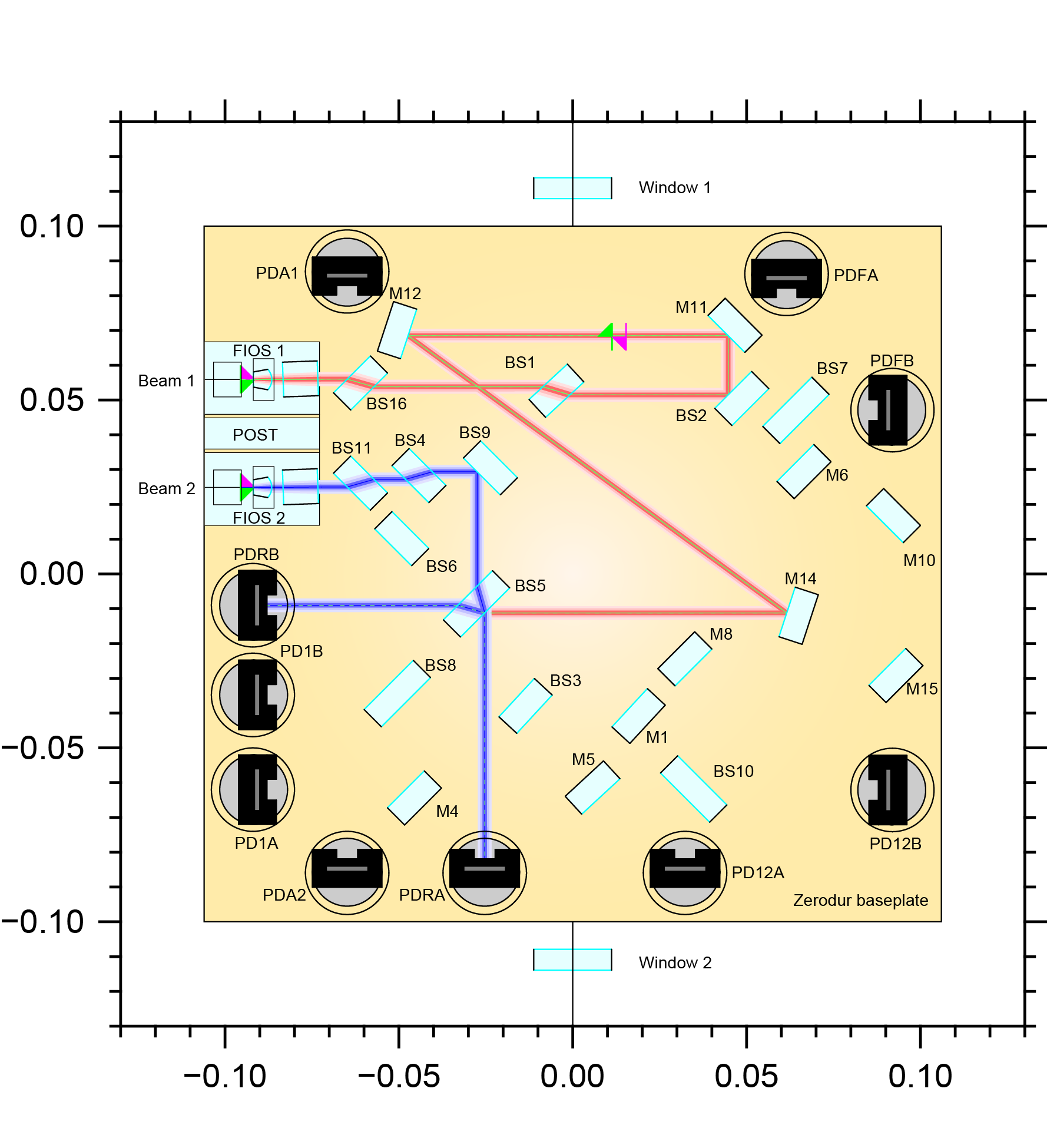}\figsubcap{c}}
	\hspace*{4pt}
	\parbox{2.1in}{\includegraphics[width=2in]{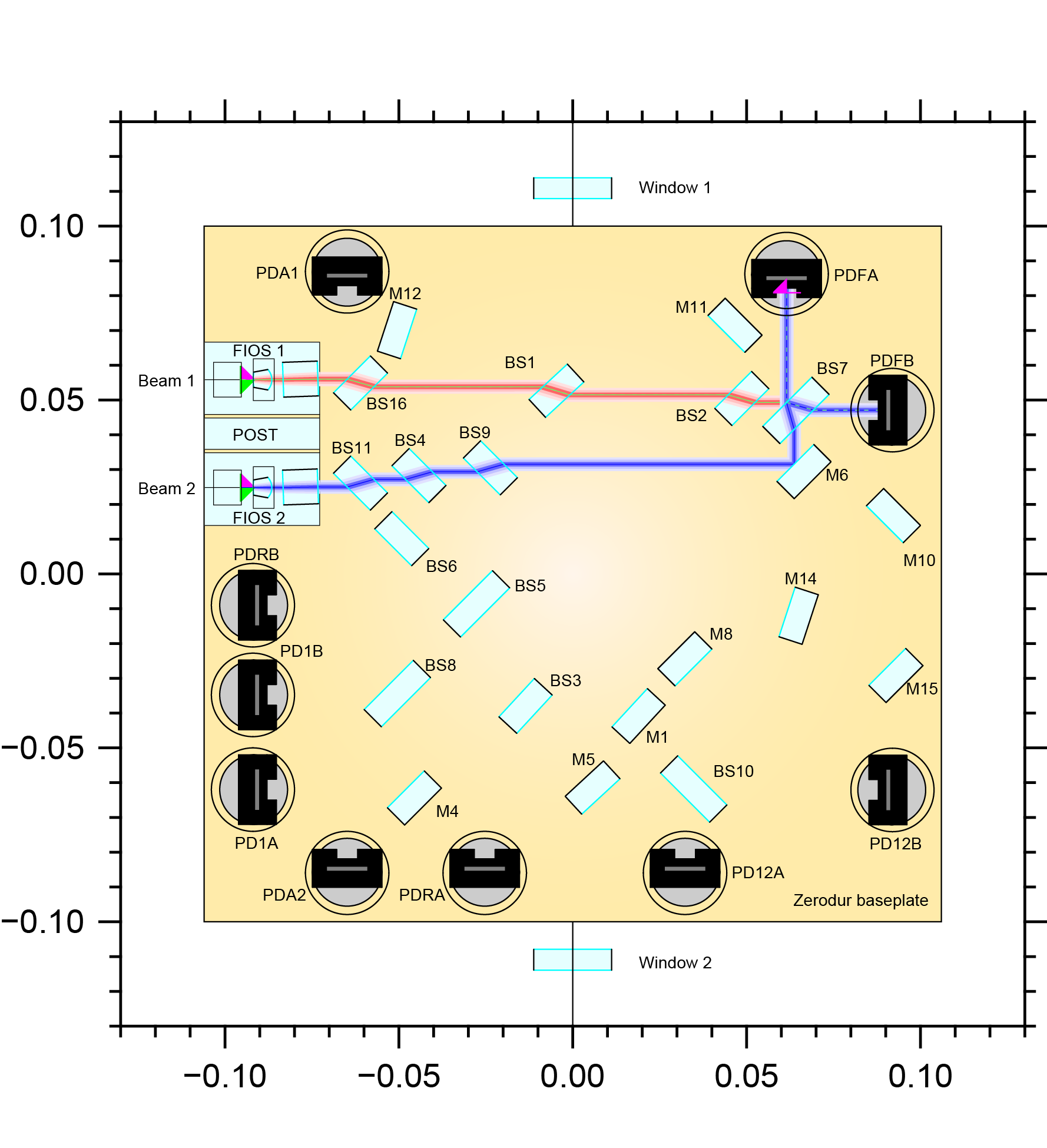}\figsubcap{d}}
	\caption{The four interferometers on the LISA Pathfinder optical bench. (a) X12 interferometer. (b) X1 interferometer. (c) Reference interferometer XR. (d) Frequency interferometer XF.}%
	\label{fig:4_IFOs}
\end{center}
\end{figure}

The main interferometer is the X12 interferometer which measures the relative distance of the two test masses. In this interferometer, the so-called measurement beam is steered with mirrors in such a way that it hits both test masses and gets reflected on the respective surfaces. The other beam, the reference beam, stays on the optical bench. If now the relative distance in between the two test masses changes, the phase of the measurement beam will reflect this and so will the beatnote signal. In each of the four interferometers, the same beatnote is recorded by two photodiodes whose signals are combined. This scheme is called balanced detection\cite{balanced_detection}. Having two photodiodes for each interferometer is not only useful in terms of redundancy but also because with this balanced detection scheme, some phase errors that arise due to relative intensity noise in the laser cancel\cite{balanced_detection}.

Similarly, the X1 interferometer measures the relative distance between the TM1 and the satellite. Here, the measurement beam hits only test mass 1 and not both test masses. The measurement beam stays on the optical bench.

There are also two auxiliary interferometers on the optical bench that are required to obtain a position measurement at the required noise level. These are the reference (XR) and the frequency interferometer (XF). In both of these, the two beams stay completely on the optical bench and do not hit any of the two test masses. To understand the reference interferometer, it is important to note that the optical path length in the fibres may fluctuate, for example due to temperature fluctuations. These fibres are not as stable as the optical bench. This noise will show up as a common mode phase noise in all four interferometers. Moreover, the two fibres for the measurement beam and the reference beam differ in optical path length by approximately \SI{38.2}{\centi\metre}\cite{threeOB}. The path of the light in the reference interferometer is chosen such the beam with the shorter fibre has the longer path on the optical bench to match the path length to minimise the coupling of laser frequency noise, as explained below. Thus, the reference interferometer measures exactly the phase noise resulting from the common mode path length noise. This measurement is subtracted on board LISA Pathfinder from the other three interferometer measurements.
In addition to the noise mitigation by subtraction in software, there is the so-called optical path length difference (OPD) control loop\cite{Control_Loops_2011}. Based on the measurement in the reference interferometer, two piezo-actuators adjust the optical path length in the laser modulator unit. For further details we also refer to \citenum{LISA_symp_2016_OPD}. This suppression mechanism was found to be necessary in test campaigns on ground due to non-linear couplings from the radio signals that drive the acousto-optic modulators which modulate the light to have two different frequencies \cite{TN3028}.

The fourth interferometer is the frequency interferometer (XF). The path of the two beams of the frequency interferometer on the optical bench is equally long, such that there is an intentional path length difference arising from different fibre lengths. This amplifies the laser frequency noise $\delta f$ following\cite{GHTN3010}
\begin{equation}
\label{eq:coupling_of_lf_with_pl_mismatch}
\delta \phi = 2\pi \frac{\Delta s }{c} \delta f\, ,
\end{equation}
with $\delta \phi$ being the resulting phase error which cannot be distinguished from the measured signal. The coupling is determined by the path length
difference, $\Delta s$, and $c$ denotes the speed of light. This measurement of the laser frequency noise is fed into a controller that commands a piezo and a heater to stabilise the laser frequency. For details, we refer to \citenum{Paczkowski2017}. 

After many years of development and testing on ground, not only for the OMS but also for all the other key systems, they were ready to be finally integrated into the satellite.
\subsection{Launch and Mission Operations}
LISA Pathfinder was launched by a  ESA-VEGA rocket by Arianespace on December, 3rd 2015 at 04:04 UTC from Kourou, French Guiana. After six apogee raising manoeuvres, there was the final burn that got LISA Pathfinder out of the elliptical orbit around the Earth and onto its way to the Lagrange point, L1. This Lagrange point is located in between Earth and the Sun, at a distance of approximately $\SI{1.5}{\giga\metre}$. 
Simply speaking, objects at this point orbit the Sun simultaneously with Earth which leads to a very small local gravity field. This makes it an ideal location for a high-precision low acceleration measurement which is what we want to do with LISA Pathfinder. However, the precise orbit of LISA Pathfinder is a $\SI{500 000}{\kilo\metre}\,\mathrm{x}\,\SI{800 000}{\kilo\metre}$ Lissajous orbit around the L1 point \cite{LPF_overview_LISA_Symp_Florida}. Already while the satellite was still on its way to L1, one unit after the next was switched on and checked for its principal functionality. This process is known as the in-orbit commissioning and it started on January, 11th 2016. After the successful in-orbit commissioning of LISA Pathfinder, the nominal mission operations phase of the LISA Technology Package (LTP) began on the first of March 2016. The nominal mission duration of three month continued until June, 26th 2016. This period was longer than June, 1st since the days where station keeping manoeuvres took place were not counted for the three months of science operations. These station-keeping manoeuvres were necessary to keep the satellite on its orbit but no residual acceleration noise measurements could be performed during these days. The nominal LTP operations phase was followed by the Disturbance Reduction System (DRS) operations. The DRS is a NASA payload on board of LISA Pathfinder. It tests a slightly different drag-free and attitude control system and a different set of $\SI{}{\micro\N}$ thrusters. This phase was followed by an LTP mission extension which lasted until the final shut down of the LISA Pathfinder satellite on July, 18th 2017.

The technology demonstrator mission LISA Pathfinder can be seen as our laboratory in space. From an operating point of view however, the daily routine is different. Due to the limited duration of the operations and the numerous measurements needed, the time had to be well organised.
Once scientists have decided for the next experiments to take place, they are inserted into a human readable schedule which takes into account the state that the satellite has to be in and the duration of the experiment. This is transformed by the Science and Technology Operations Centre (STOC) into a series of pre-defined command-blocks. The Mission Operations Centre (MOC) then controls the expansion of the command-blocks into telecommands. These telecommands are then uploaded to the satellite for execution. The resulting data is transferred to Earth with a limited data rate. This is why, in contrast to many other experiments, the sampling frequency of many LISA Pathfinder science data channels was $\SI{1}{\Hz}$ or $\SI{10}{\Hz}$. The data was transferred to the servers and stored as so-called analysis objects in LTPDA\cite{LTPDAReference}, a dedicated MATLAB$^\text{\copyright}$ toolbox developed for LISA Pathfinder. A major advantage of this software is that it tracks the history of the data used through all the stages of the analysis that is done to it. The data analysis was performed during the nominal operations phase in close to real time. That means, a team of engineers and scientists was located at the European Space Operations Centre in Darmstadt, Germany to analyse the arriving data. On the basis of these results, the next experiments were chosen. 

In the following section, we will describe selected measurements and results obtained with the procedure and software explained in this section.

\section{The Physics of LISA Pathfinder}
\subsection{The residual acceleration $\Delta g$ explained}
\label{subsec:delta_g_explained}
As mentioned already, the main measurement on LISA Pathfinder was the measurement of the residual acceleration noise. For this measurement, we take the second derivative of the measured relative position of the two test masses, denoted as $x_{12}$, and subtract the estimated applied forces on TM2, $g_{\mathrm{c}}(t)$:  
\begin{equation}
	\Delta g = \ddot{x}_{12} - g_{\mathrm{c}}(t) \,.
	\label{eq:delta_g_preliminary}
\end{equation}
All quantities in this equation are always given per unit mass in this document and thus forces have the unit of acceleration, too.
Equation \ref{eq:delta_g_preliminary}, however, is only the starting point for the estimation of the residual acceleration and is not complete yet. In addition, we have to take known cross-couplings into account. Let us note that Equation \ref{eq:delta_g_preliminary} is simplified in the sense that it assumes the effect of the control force to be immediate. In reality, the DFACS control loops have a non-zero delay which is neglected here for simplicity.

Two long-known terms arise from the motion of the test masses in force gradients. This coupling is called a stiffness $\omega^2$ and it acts like a spring with a negative spring constant. That means if a test mass gets accelerated into one direction, the resulting motion will be enhanced. In contrast, the usual springs have a positive spring constant and pull back the deflected object. The stiffness of each test mass is the sum of two stiffness contributions: a gravitational contribution and an electrostatic contribution. The gravitational component is due to the fact that the satellite is balanced in such a way that the test masses are subject to a minimal gravitational force at their optimal positions. Even a slight deviation from this position causes the test masses to be pulled to the centre of mass of the satellite which leads even further away from the optimal position. The stiffness also has an electrostatic component because there is a change in the electrostatic force acting on the test mass due to a position change of the test mass. The amplitude of this coupling depends on the so-called actuation authority. This is the maximum amount of actuation forces or torques that is permitted in the current setting on the satellite. From this number follows the amplitude of the $\SI{100}{\kilo\Hz}$ AC voltage that is used instead of a noisy DC position readout. And the electrostatic component of the stiffness depends on exactly these voltages. Preliminary results indicate the gravitational contribution is larger than the electrostatic\cite{sysID_yak}. 

Including now the stiffness, the acceleration of each of the test masses reads
\begin{align}
\ddot{x}_1 &= g_1 - \omega_1^2\left( x_1 -x_{\mathrm{SC}} \right) \\
\ddot{x}_2 &= g_2 - \omega_2^2\left( x_2 -x_{\mathrm{SC}} \right) + g_{\mathrm{c}}(t) 
\end{align}
Combining these two equations to obtain the residual differential acceleration yields:
\begin{align}
	\Delta g &= g_2 - g_1 \, ,\\
	             &= - \ddot{x}_1 - \omega_1^2\left( x_1 -x_{\mathrm{SC}} \right) + \ddot{x}_2 + \omega_2^2\left( x_2 -x_{\mathrm{SC}} \right) - g_{\mathrm{c}}(t) \,.
\end{align}
Next, we take into account that the interferometer measures $o_{12} = x_2 - x_1$ and replace $x_2$ with $x_2 = o_{12} +x_1$ accordingly to obtain:
\begin{equation}
		\Delta g = \ddot{o}_{12} -  \omega_1^2\left( x_1 -x_{\mathrm{SC}} \right)  +  \omega_2^2\left( o_{12}+ x_1 -x_{\mathrm{SC}} \right) - g_{\mathrm{c}}(t) \,.
\end{equation} 
This equation is also simplified for in reality, the interferometer is a system with readout noise and possible cross-sensing from other degrees of freedom or measurement channels. The coupling from other degrees of freedom will be partially included later in the derivation of $\Delta g$ in the cross-coupling term. Leakage from other OMS measurement channels is not considered in this summary and we refer the reader to \citenum{vetrugno_karnesis_2017}.
Finally, the X1 interferometer measures in fact the position of TM1 with respect to the satellite such that we can redefine $o_1 := x_1 - x_{\mathrm{SC}}$. Introducing also the differential stiffness $\Delta \omega = \omega_2^2 -\omega_1^2$ allows us to write:
\begin{equation}
	\Delta g = \ddot{o}_{12} +  \omega_2^2  o_{12} + \Delta \omega o_1 -g_c \,.
\end{equation}
From this equation, it can be seen that as the two test masses react slightly differently to the motion of the satellite (as measured by the X1 interferometer) due to the two different stiffnesses; this acceleration that originates in the satellite looks like a differential acceleration. The precise values of these stiffnesses have been estimated via dedicated experiments. These experiments consist of signal injection into the drag-free and the suspension loop. For details, we refer to \citenum{vetrugno_karnesis_2017}.

However, during the mission, three more effects that need to be taken into account have been identified. One of them is the interferometer pick-up of satellite motion $g_{\mathrm{cross talk}}(t)$, as explained in \citenum{Wanner_LISA_symp}. In addition, the centrifugal forces $g_{\Omega}(t)$\cite{2016PhRvL.116w1101A} and other spacecraft angular acceleration effects $g_{\mathrm{decorr}}(t)$, which will be explained in a future publication, have to be subtracted.
The equation for the residual differential acceleration currently reads, therefore,
\begin{equation}
	\Delta g = \ddot{o}_{12} +  \omega_2^2  o_{12} + \Delta \omega o_1 -g_c -g_{\Omega}(t) - g_{\mathrm{cross talk}}(t) - g_{\mathrm{decorr}}(t) \,.
\end{equation}
Neglecting the spacecraft angular acceleration effects, which do not provide a significant contribution in April 2016, when the data was taken, the measured residual acceleration was found to be 
$\SI[per-mode=reciprocal]{5.2 \pm 0.1}{\femto\metre\per\second\squared\per\sqrthz}$ in the frequency range from $\SI{0.7}{\milli\Hz}$ to $\SI{20}{\milli\Hz}$ \cite{2016PhRvL.116w1101A}. Towards lower frequencies, the noise increases but remains below $\SI[per-mode=reciprocal]{12}{\femto\metre\per\second\squared\per\sqrthz}$ down to $\SI{0.1}{\milli\Hz}$. The reasons for this increase are not identified yet but laser radiation pressure fluctuations as well as thermal gradient and magnetics force effects are not limiting the performance in this frequency range. At frequencies above $\SI{60}{\milli\Hz}$, the interferometer readout noise of $\SI[per-mode=reciprocal]{34.8 \pm 0.3}{\femto\mnoise}$ is dominating\cite{2016PhRvL.116w1101A}. 

These results have exceeded even the most optimistic expectations. These measurements show that LISA Pathfinder was the quietest place in the universe ever measured. The residual acceleration level is more than five times better than originally required. In this frequency range, the LISA Pathfinder results are even close, that is within a factor of $1.25$, to the required LISA performance. In addition, the interferometer noise performance is more than one hundred times better than expected from ground test campaigns. This is very encouraging for the further development and construction of LISA for it shows that indeed, the technology to achieve the required levels of free-fall is available and working. This is underlined from an operational point of view by the fact that during the first 
55 days of operations, already more than 650 hours of residual acceleration noise measurements could be taken\cite{2016PhRvL.116w1101A}.

In view of LISA, it is important to not only measure the residual acceleration noise level on LISA Pathfinder but to make the best use of this unique opportunity of such a laboratory in space to understand the system and the different subsystems in as much detail as possible. 
\subsection{The GRS measurements}
\label{subsec:GRS_measurements_and_dm}
\begin{figure}
	\centering
	\includegraphics[width=0.45\textwidth]{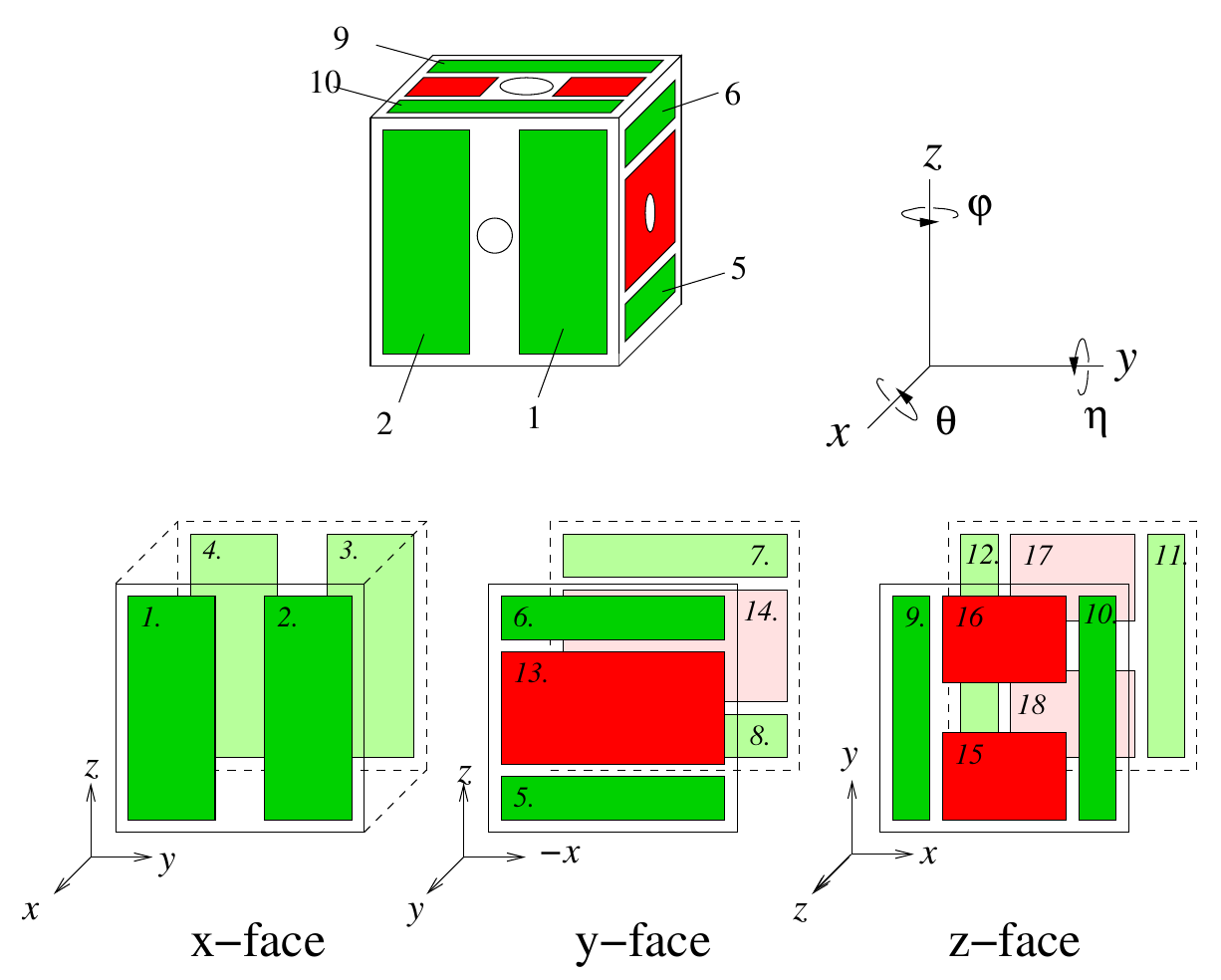}
	\caption{The GRS layout with the sensing and actuation electrodes in green and the injection electrodes, which apply the $\SI{100}{\kilo\Hz}$ voltage, in red. Reprint from \citenum{S2-SRO-TN-2001-IS}.}
	\label{fig:GRS_electrodes_pic}
\end{figure}

One example for detailed analysis is the electrostatic sensing and actuation system which is part of the GRS, as explained above. This system consists of a set of 18 electrodes as shown in Figure \ref{fig:GRS_electrodes_pic}. The electrodes in green are used for sensing the position of the test masses and, at the same time, to apply the commanded forces as determined by the DFACS. The electrodes in red are used to apply the AC $\SI{100}{\kilo\Hz}$ voltages to the test masses. They polarise the test masses and in fact, the position measurements are the result of the demodulation of the corresponding difference in current in a capacitive-inductive bridge, as depicted for example in \citenum{Neda_GRS}.  
With this system of electrodes and the corresponding circuitry, it is possible to measure all six degrees of freedom for each of the two test masses. These measurements are, however, noisier than the measurements of the OMS, for the degrees of freedom for which the later provides measurements. For example, with the GRS, the noise in the position sensing is below $\SI[per-mode=reciprocal]{2.4}{\nano\mnoise}$ in the frequency range from $\SI{1}{\milli\Hz}$ to $\SI{0.5}{\Hz}$ \cite{capacitive_sensing_results}. With the OMS, the noise in the position sensing along the sensitive $x$-axis is at the level of $\SI[per-mode=reciprocal]{34.8 \pm 0.3}{\femto\mnoise}$ only and therefore was found to dominate $\Delta g$ for frequencies above $\SI{60}{\milli\Hz}$. When comparing the precision of the two sensing systems, we find that the OMS is roughly five orders of magnitude less noisy. A very important characteristic of the set-up shown in Figure \ref{fig:GRS_electrodes_pic} is the size of the gaps between each of the test masses and the respective electrodes. Along the sensitive $x$-axis, the gap is $\SI{4}{\milli\metre}$, along the $y$-axis $\SI{2.9}{\milli\metre}$ have been chosen and $\SI{3.5}{\milli\metre}$ for the $z$-axis\cite{S2-SRO-TN-2001-IS}. These are larger than all other sensing gaps implemented on drag-free space missions so far \cite{capacitive_sensing_results}.
In nominal science operations, actuation is necessary to keep the two test masses at their nominal position, with the exception of the $x$-degree of freedom of the free-falling TM1, and to avoid tilts of the test masses along all rotational degrees of freedom. The maximum actuation force that could be applied during nominal science operations is the limit of the so-called high resolution mode of the actuation system, which is always switched on during nominal science operation, and whose maximum is $\approx \SI{2.2}{\nano\newton}$\cite{S2_ASD_TN_2011}. The size of these actuation forces and torques strongly depends on the size of the local gravity field on board the satellite. That means, the better the balancing of the masses of all satellite components, the smaller the remaining static gravitational force, $\Delta g_{\mathrm{DC}}$, that has to be compensated. Prior to launch, the satellite had to be designed such that a local gravitational acceleration of $\Delta g_{\mathrm{DC}} = \SI[per-mode=reciprocal]{650}{\pico\metre\per\second\squared}$ would not be exceeded along the sensitive axis \cite{LPF_self_gravity}. In flight, the same quantity was found to be below $ \SI[per-mode=reciprocal]{50}{\pico\metre\per\second\squared}$ \cite{2016PhRvL.116w1101A}. 
The maximum force or torque that can be applied by the actuation system in the current setting, the so called actuation authority, is a very important quantity because it determines the level of stray force noise that is caused by the actuation. This can be understood by noting that the actuation authority determines the amplitude of the carrier voltages that are applied to the electrodes. It is by modulating these carrier voltages that the electrostatic forces are applied to the test masses. However, a certain level of fluctuations cannot be avoided. On ground, the relative amplitude stability of the actuation voltages of the flight amplifiers was found to be between $ 3 $ and $\SI{8}{\text{ppm}\per\sqrthz}$ \cite{LPF_self_gravity}.
In summary, this means the higher the necessary maximum voltages, the noisier the actuation system and the more undesired force noise is experienced by the test mass. Consequently, the very low static local gravity field allowed us to use the actuation system in an extremely low noise configuration. This enabled us to make the very quiet free-fall measurements as shown in \citenum{2016PhRvL.116w1101A}. If a higher actuation authority were necessary, the noise in the frequency range from $\SI{0.1}{\milli\Hz}$ to approximately $\SI{3}{\milli\Hz}$ would have been higher.
To understand if there are other effects in the electrostatic actuation system resulting from actuation noise which is independent of the actuation authority, an experiment called the drift mode or free-flight experiment is used. In this experiment, the continuous control of TM2 along $x$ is replaced with a control scheme based on short force impulses which are followed by several minutes of uncontrolled drift of TM2 along the $x$ direction. For details of the implementation of this experiment, we refer to \citenum{Grynagier2010} and \citenum{LPF_DM_Grynagier}. The results of the test of this experiment on Earth at the torsion pendulum facility at the University of Trento can be found in \citenum{LPF_FF_ground_test}. The interesting data analysis approaches and first results can be found in \citenum{FF_LISA_Symposium_2014} and \citenum{FF_LISA_Symposium_2016}.

Another example for detailed studies is the Optical Metrology System. Preliminary results indicate that the main measurement of the relative position of the two test masses is more than a hundred times less noisy than the tests performed on ground. This applies to frequencies above approximately $\SI{0.4}{\Hz}$. Below this frequency, the interferometer signal is dominated by the pick-up of test mass and satellite motion. Towards even lower frequencies, the motion of the test mass itself is much larger than the noise. The minimal level of the white OMS noise is determined by the noise in the phasemeter \cite{2016PhRvL.116w1101A}. In addition, possible contributions from laser frequency noise \cite{Paczkowski2017} and relative intensity noise \cite{RIN_LISA_Symp_2016} are under investigation. 
\subsection{Charge-related measurements and results}
In addition, another set of detailed investigations has been performed to understand the influence of charge on LISA Pathfinder. There are two ways in which charge can produce an undesired force noise. One of them is the mixing of a fluctuating charge with stray DC potentials around the test masses. A certain level of charge fluctuations cannot be avoided for they result from high-energy cosmic rays and solar energetic particles which hit the TMs and charge them. Consequently, the remedy is to compensate the stray potentials as much as possible. This is done by adding small $\SI{}{\milli\V}$ DC voltages to the electrodes. The second mechanism is the mixing of a static charge with noisy stray voltages. These stray voltages can be caused by surface patch potentials and noise in the GRS electronics. The influence of these stray voltages is minimised by the comparatively large gap size, compare Section \ref{subsec:GRS_measurements_and_dm}. In addition, both test masses are discharged via the photoelectric effect using UV-light. Accordingly, two kinds of charge-related experiments have been performed. The charge itself was estimated by applying quasi DC modulations of the voltages to the surrounding electrodes and measuring the resulting force. Vice versa, the stray potentials have been estimated based on the resulting force from a deliberate change in charge. In summary, it can be concluded that charge-related effects are far from limiting the LISA Pathfinder performance \cite{LPF_charge_results}.

These are only selected examples of detailed system studies and more results will be published soon.
\section{The future of LISA Pathfinder: LISA}
While the first proposal of LISA to ESA and NASA, at that time involving four satellites, dates back to the 1993, we will summarise only the recent development of LISA here. 
In March 2013, the European Space Agency called for science themes which could be addressed by future missions of their large scale class. This call was answered by many proposals, one of them being `The Gravitational Universe' \cite{the_gravitational_universe}. In November 2013, this science theme was selected for the L3 mission slot with a planned launch in 2034. In October 2016, ESA called for mission concepts which implement the science theme  `The Gravitational Universe'. LISA was proposed as such a mission concept in January 2017\cite{LISA_L3} and selected in June 2017. At the time of writing, the development of LISA is continuing at a high speed and with great enthusiasm.

\section{Conclusions}
LISA Pathfinder is the technology demonstrator mission for the future gravitational wave observatory, LISA. The satellite was launched on December, 3rd 2015 and operated successfully until the final shut down on July, 18 2017. It was not only shown that it is possible to have a free-falling test mass in space whose residual acceleration due to spurious forces is below the required level of $\SI[per-mode=reciprocal]{30}{\femto\metre\per\second\squared\per\sqrthz
}$ for $\SI{}{\milli\Hz}$ frequencies but also that this requirement is fulfilled with a large margin. This means, we are already approaching the necessary free-fall levels for LISA. In addition, from an organisational point of view, it was shown that such a satellite, in which the payload is the science instrument, can be operated in close cooperation between the European Space Agency, industry partners and scientists. Important lessons learned during all of the mission phases from hardware development to mission operations are currently being summarised and will provide additional input for the development of LISA. We leave the details to future publications. 

The performance of the Optical Metrology System with a readout noise of only $\SI[per-mode=reciprocal]{34.8 \pm 0.3}{\femto\mnoise}$ \cite{2016PhRvL.116w1101A} is more than one hundred times better than the latest measurement of the test campaigns on Earth. In other words: we were not only able to operate the first public laser interferometer system in space but also to measure $\SI{}{\femto\metre}$ instead of the required $\SI{}{\pico\metre}$ accuracy. This is a very successful test of the local interferometry system on board of each LISA satellite. For the inter-satellite interferometry, the noise level is expected to be above the $\SI{}{\femto\metre}$ level. In general, the satellite and all key subsystems to achieve free-fall, as for example the OMS, the GRS and the charge management, could be characterised in detail with dedicated experiments. A deep understanding of each subsystem is very useful for the remaining development of LISA.

LISA has been selected as a Mission Concept in June 2017 and currently work is well under way to clarify the remaining science questions and to start the industrial engineering. Even though the launch date in 2034 seems in the far future and many key components have shown excellent performance on LISA Pathfinder, some development concerning for example the inter-satellite interferometry and the required telescopes as well as the link in between the two optical benches on one satellite is still required. Even though some aspects of inter-satellite interferometry will be tested on the GRACE Follow-On mission\cite{LRI}, the future interferometry development for LISA will include more dedicated hardware testing. It necessarily takes a certain time as, for example, complex optical benches need to be built. That is the reason why the development has already started at the time of writing.

To conclude, LISA Pathfinder has exceeded all expectations and shown that LISA is feasible. 
\section*{Acknowledgments}
This work has been made possible by the LISA Pathfinder mission, which is part of the space-science
program of the European Space Agency. The French
contribution has been supported by CNES (Accord Specific de Projet No. CNES 1316634/CNRS 103747), the CNRS, the Observatoire de Paris and the University Paris-Diderot. E.P. and H.I. would also like to acknowledge the financial support of the UnivEarthS Labex program at Sorbonne Paris Cit\'e (Grants No. ANR-10-LABX-0023 and No. ANR-11-IDEX-0005-02). The Albert-Einstein-Institut acknowledges the support of the German Space Agency, DLR. The work is supported by the Federal Ministry for Economic Affairs and Energy based on a resolution of the German Bundestag (Grants No. FKZ 50OQ0501 and No. FKZ 50OQ1601). The Italian contribution has been supported by Agenzia Spaziale
Italiana and Instituto Nazionale di Fisica Nucleare. The Spanish contribution has been supported by Contracts No. AYA2010-15709 (MICINN), No. ESP2013-47637-P, and No. ESP2015-67234-P (MINECO). M.N. acknowledges support from Fundacion General CSIC Programa ComFuturo). F.R. acknowledges support from a Formacin de Personal Investigador (MINECO) contract. The Swiss contribution acknowledges the support of the Swiss Space Office (SSO) via the PRODEX Programme of ESA. L.F. acknowledges the support of the Swiss National Science Foundation. The UK groups wish
to acknowledge support from the United Kingdom Space Agency (UKSA), the University of Glasgow, the University of Birmingham, Imperial College London, and the
Scottish Universities Physics Alliance (SUPA). J.I.T. and J.S. acknowledge the support of the U.S. National Aeronautics and Space Administration (NASA).

\bibliographystyle{ws-procs961x669}
\bibliography{my_sources}
\end{document}